\documentclass[12pt]{iopart}
\usepackage{iopams}  
\usepackage{graphicx}
\def\be{\begin{equation}}
\def\ee{\end{equation}}

\newcommand{\beq}{\begin{equation}}
\newcommand{\eeq}{\end{equation}}
\newcommand{\ber}{\begin{eqnarray}}
\newcommand{\eer}{\end{eqnarray}}
\newcommand{\berr}{\begin{eqnarray*}}
\newcommand{\eerr}{\end{eqnarray*}}

\begin{document}

\title{ 
Quarkonium in Hot Medium
}

\author{P\'eter Petreczky 
}
\address{ 
Department of Physics and RIKEN-BNL Research Center, Brookhaven
National Laboratory, Upton, New York, 11973
}

\begin{abstract}
I review recent progress in studying quarkonium properties
in hot medium as well as possible consequences for quarkonium
production in heavy ion collisions.
\end{abstract}

\pacs{11.15.Ha, 11.10.Wx, 12.38.Mh, 25.75.Nq}

%\submitto{\JPG}

%\maketitle

%
\section{Introduction}
\label{intro}

There has been considerable interest in studying quarkonia in hot
medium since the publication of the famous Matsui and Satz paper 
\cite{Matsui:1986dk}. It has been argued that color screening in a deconfined QCD 
medium will suppress the existence of quarkonium states, signaling the 
formation of a quark-gluon plasma (QGP) in heavy-ion collisions. Although 
this idea was proposed a long time ago, first principle QCD calculations, 
which go beyond qualitative arguments, have been performed only recently. 
Such calculations include lattice QCD determinations of quarkonium 
correlators 
\cite{Umeda:2002vr,Asakawa:2003re,Datta:2003ww,Jakovac:2006sf,Aarts:2007pk}; 
potential model calculations 
of the quarkonium spectral functions with potentials based on lattice QCD 
\cite{Digal:2001ue,Wong:2004zr,Mocsy:2005qw,Alberico:2006vw,Cabrera:2006wh,Mocsy:2007yj,Mocsy:2007jz}
(see also Ref. \cite{satz_rev} for a review)),  
as well as effective 
field theory approaches that justify potential models and reveal new medium 
effects 
\cite{Laine:2006ns,Laine:2007qy,Laine:2007gj,Laine:2008cf,Brambilla:2008cx}.  
Spectral properties of heavy quark bound states are important ingredients
in modeling of heavy quarkonium production in hot medium as will be discussed later.

%%%%%%%%%%%%%%%%%%%%%%%
\section{Color screening and deconfinement}

At high temperatures, strongly-interacting matter undergoes a deconfining
transition to a quark-gluon plasma (QGP). 
This transition is triggered by a rapid increase of the energy and
entropy densities, as well as the disappearance of hadronic states. (For  
recent reviews, see Ref.~\cite{DeTar:2009ef,petr_qm09,fodor_rev}). 
According to current lattice calculations, at zero net baryon density 
deconfinement occurs at $T_c \sim (170-195)~$MeV \cite{hoteos,oureos}
\footnote{ 
In the view of crossover nature of the deconfinement transition it is
difficult to define the corresponding temperature interval precisely.
Moreover due to discretization errors there is a discrepancy in the value
of the deconfinement transition temperature obtained in calculations using
different discretization schemes. Calculations with so-called
stout staggered fermion action give a deconfinement transition temperature
around $170$MeV \cite{fodor06,fodor09}. 
The new calculation with so-called HISQ action \cite{hisq} seem to support
this scenario.}

The QGP is characterized by color screening: the range of interaction between
heavy quarks becomes inversely proportional to the temperature. Thus at
sufficiently high temperatures, forming a bound state with a heavy quark ($c$ 
or $b$) and its anti-quark becomes impossible. 
\vspace*{-0.5cm}
\begin{figure}[htbp]
\begin{center}
\includegraphics[width=7.7cm]{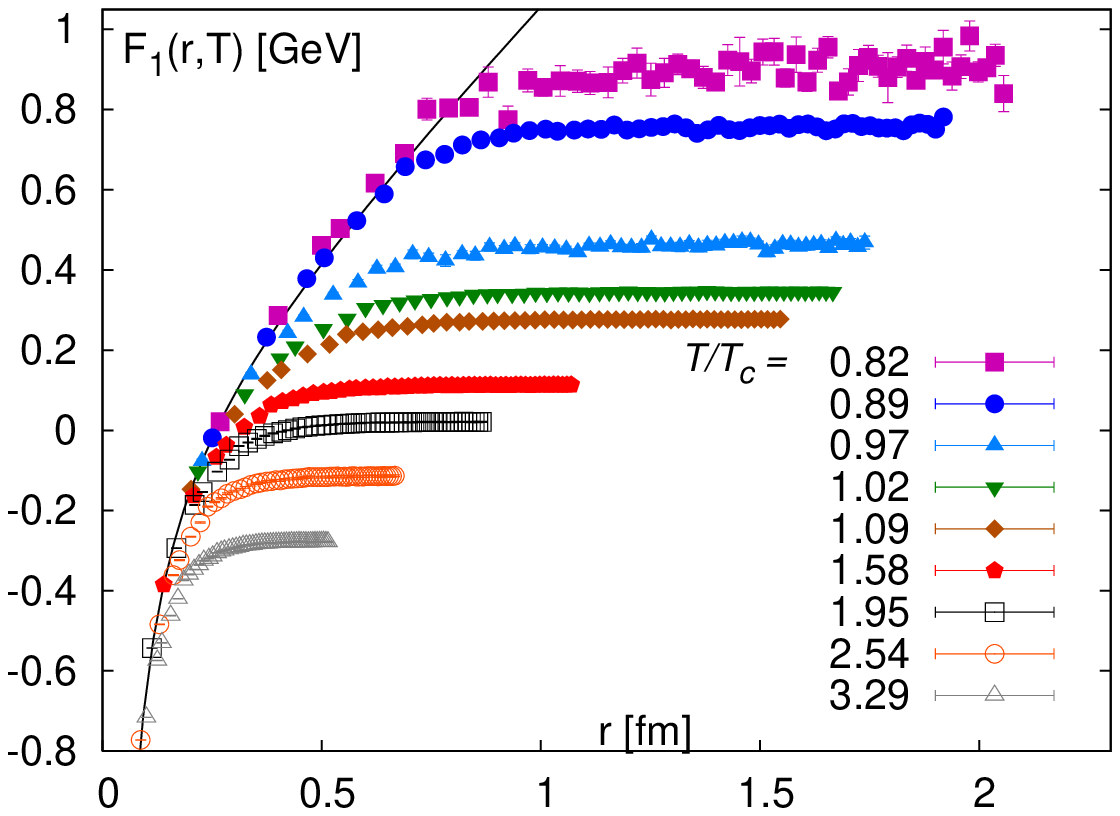}
\includegraphics[width=7.7cm]{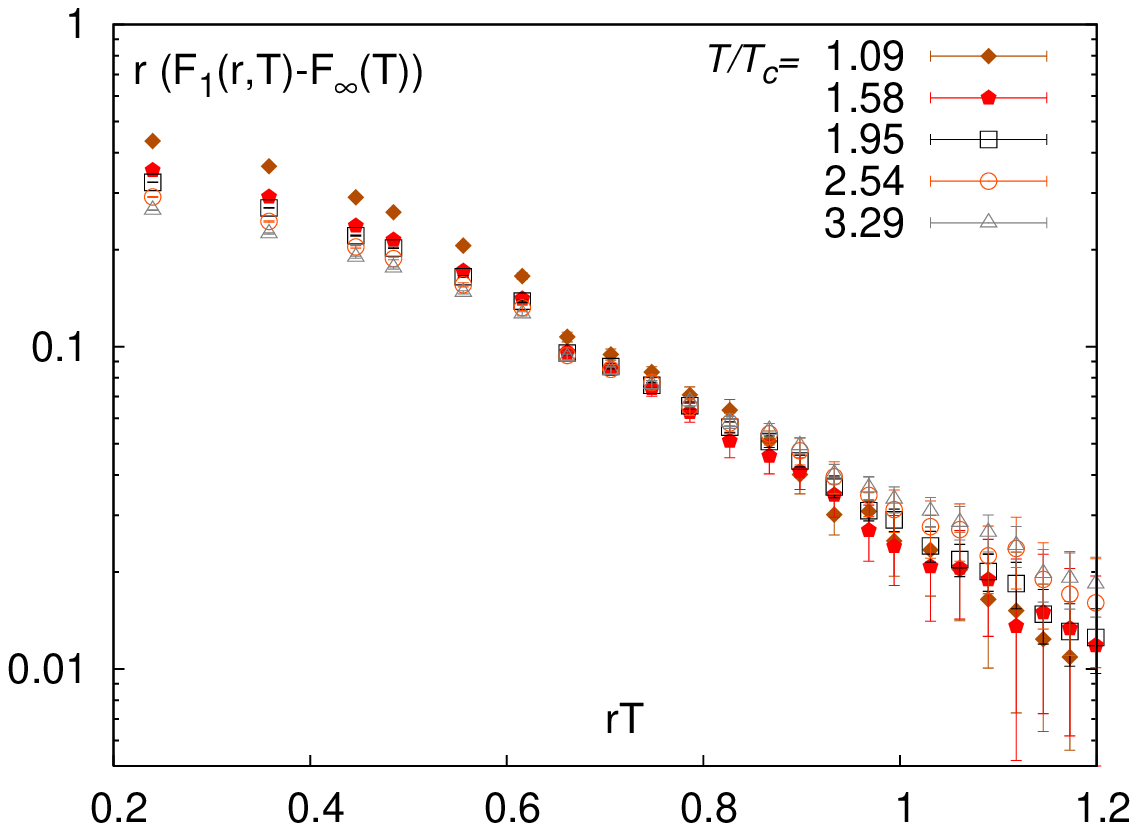}
\vspace*{-0.5cm}
\caption[]{Static quark singlet free energy versus quark separation 
calculated in 2+1 flavor QCD
on $16^3 \times 4$ lattices at different temperatures \cite{future} (right). 
The combination $r (F_1(r,T)-F_{\infty}(T))$
as function of $r T$ (left). 
The solid black line on the left plot is the
parametrization of the zero temperature potential calculated in Ref. \cite{oureos}.
}
\label{fig:f1}
\end{center}
\end{figure}
\vspace*{-0.5cm}
\begin{figure}[htbp]
\begin{center}
\includegraphics[width=7.7cm]{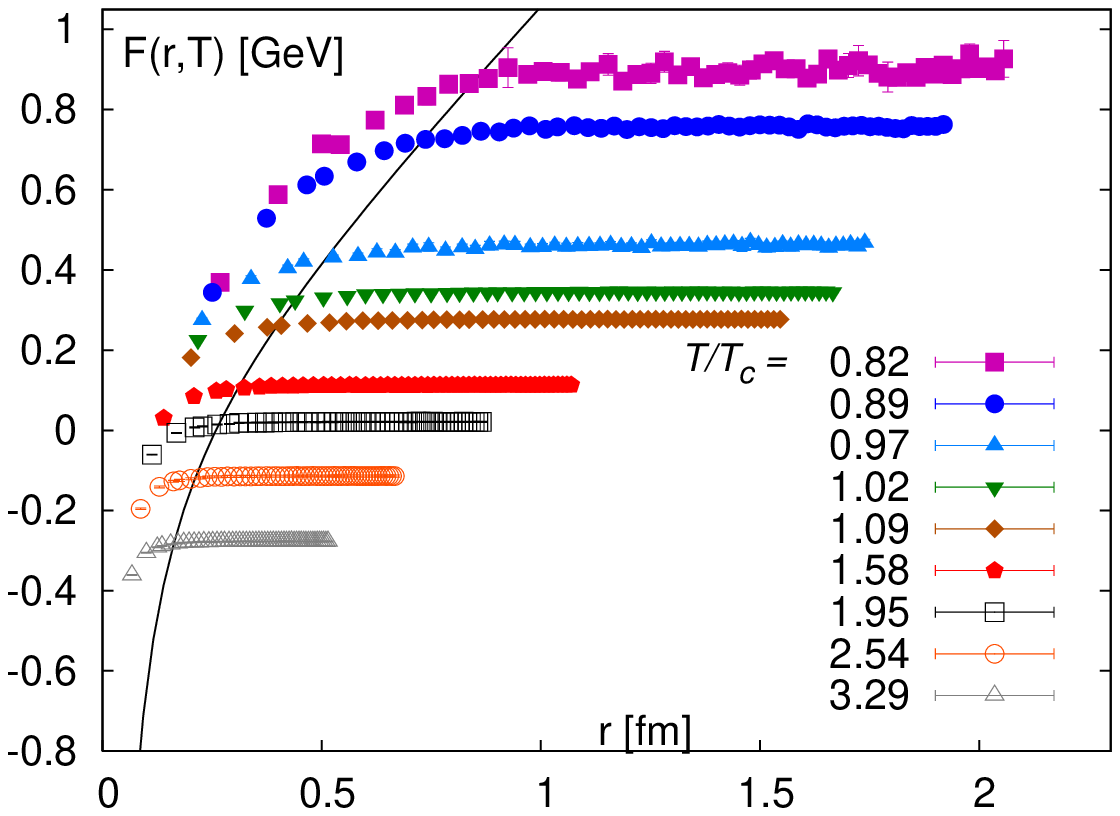}
\includegraphics[width=7.7cm]{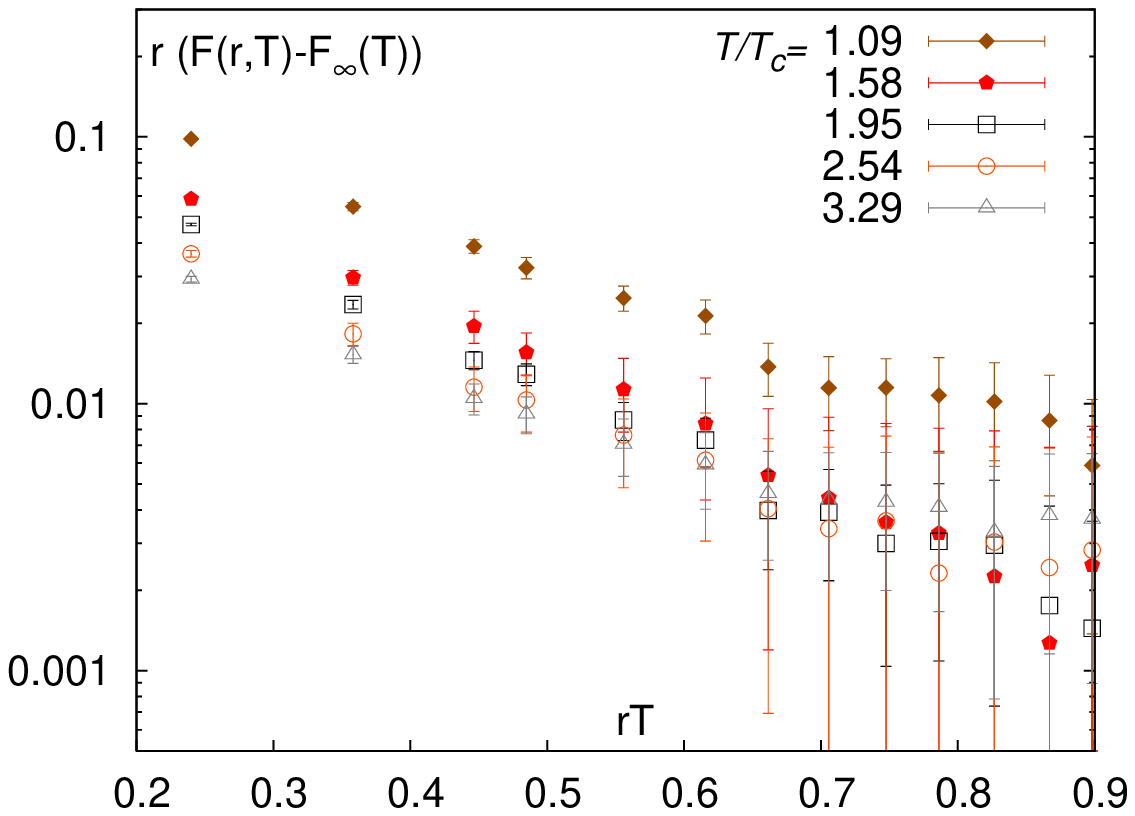}
\vspace*{-0.5cm}
\caption[]{Static quark  free energy versus quark separation 
calculated in 2+1 flavor QCD
on $16^3 \times 4$ lattices at different temperatures \cite{future} (right). 
The combination $r (F(r,T)-F_{\infty}(T))$
as function of $r T$ (left). The solid black line on the left plot is the
parametrization of the zero temperature potential calculated in Ref. \cite{oureos}.
}
\label{fig:fa}
\end{center}
\end{figure}
\vspace*{-0.7cm}
Color screening is studied on the lattice by 
calculating the spatial correlation function of a static quark and
anti-quark, which propagates in Euclidean time 
from $\tau=0$ to $\tau=1/T$ where $T$ is the temperature 
(see Ref.~\cite{Bazavov:2009us} for a recent review). 
Two types of correlation functions are usually calculated on the lattice.
The correlation function of Polyakov loops, which is also called the color
averaged correlator
\begin{equation}
G(r,T)=\frac{1}{9} \langle {\rm Tr} L(r) {\rm Tr} L^{\dagger}(0) \rangle,
\label{avg}
\end{equation}
where the temporal Wilson line is defined in terms of link variable $U_0(x_0,r)$ as 
$L(r)=\prod_{x_0=0}^{N_{\tau}-1} U_0(x_0,r)$,
and the color singlet correlator
\begin{equation}
G_1(r,T)=\frac{1}{3} \langle {\rm Tr} L(r)  L^{\dagger}(0) \rangle
\label{sin}
\end{equation}
The later is defined in Coulomb gauge \cite{okacz02,okacz04,petrov04,okacz05} 
or in a gauge invariant but path dependent manner \cite{Bazavov:2009us}. 
The singlet correlators recently have been also calculated in perturbation theory \cite{laine09}.
The free energy of static quark anti-quark
pair is defined through the logarithm of the color averaged correlator $F(r,T)=-T \ln G(r,T)$.
Analogously, the so-called singlet free energy is defined as $F_1(r,T)/T=-\ln G_1(r,T)$
\footnote{The identification of the correlators defined by Eqs. (\ref{avg}) and
(\ref{sin}) with color averaged and color singlet channels is based on perturbative arguments,
e.g. see discussion in Ref. \cite{petr_hp04}. Non-perturbatively such identification
becomes problematic as was shown in Ref. \cite{jahn04}. At short distances effective
field theory concepts can be invoked to define color singlet and color averaged channels \cite{plc_future}.}.
The singlet free energy in 2+1 flavor QCD is shown in Fig.~\ref{fig:f1} (right). 
The numerical results were obtained on $16^3 \times 4$ lattice using
p4 action and quark masses which correspond to pion mass of about $220$MeV \cite{future}.
As expected, in the zero temperature limit the
singlet free energy coincides with the zero temperature potential calculated on the lattice \cite{oureos}.
Figure~\ref{fig:f1} also illustrates that,
at sufficiently short distances, the singlet free energy is
temperature independent and equal to the zero temperature potential.
At large distances it approaches a constant value $F_{\infty}(T)$, which is
twice the free energy of an isolated static quark.  
The range of interaction decreases with increasing temperatures.  For 
temperatures above the transition temperature, $T_c$, the heavy quark 
interaction range becomes
comparable to the charmonium radius. In particular, 
the combination $r (F_1(r,T)-F_{\infty}(T))$ is expected to decay exponentially at large
distances, $r m_D>1$, with $m_D$ being the Debye mass. Indeed, 
for distances $r T>0.8$fm
the screening is exponential as can be seen from Fig. \ref{fig:f1} (right).
Based on this general observation, 
one would expect that the charmonium
states, as well as the excited bottomonium states, do not exist above the
deconfinement transition. (In the literature, this is often referred to as 
dissociation or melting). 

The path/gauge dependence of singlet free energy makes its interpretation 
somewhat complicated. Therefore, on the lattice 
one also calculates the Polyakov loop correlator. The results of the calculations
in 2+1 flavor QCD with p4 action on $16^3 \times 4$ lattices are shown in Fig. \ref{fig:fa}
in terms of the logarithm of the correlation function, i.e. the physical free energy
of static quark anti-quark pair. 
Compared to the singlet free energy the physical free energy shows much stronger temperature
dependence and is different from the zero temperature potential at all temperatures.
In the low temperature region the strong temperature dependence of the free energy can be understood 
as resulting from
significant contribution from higher excited states \cite{baza08}. 
At high temperatures, the origin strong temperature
dependence is more difficult to understand. In the short distance limit 
within the effective field theory framework it can
be attributed to the contribution from octet degrees of freedom \cite{plc_future}.
As one
can also see from Fig. \ref{fig:fa} the free energy of static $Q \bar Q$ pair 
also shows exponential screening at large distances.
Thus the strong color screening is a real effect.
In summary, lattice calculations of static quark anti-quark correlators provide evidence
that in the deconfined phase there are strong screening effect at distances relevant for
quarkonium physics. In the next section the implication of these findings on quarkonium 
spectral functions will be discussed.

%%%%%%%%%%%%%%%%%%%%
\section{Quarkonium spectral functions}

In-medium quarkonium properties are encoded in the corresponding 
spectral functions, as are their dissolution
at high temperatures. Spectral functions are defined as
the imaginary part of the retarded correlation function of quarkonium
operators. Bound states appear as peaks in the spectral functions.
The peaks broaden and eventually disappear with
increasing temperature. The disappearance of a peak signals the melting of 
the given quarkonium state. 

In lattice QCD, the meson
correlation functions, $G(\tau,T)$, are calculated in Euclidean time. 
These correlation functions are related 
to the spectral functions $\sigma(\omega,T)$ as
\begin{equation}
G(\tau,T)=\int_0^{\infty} d \omega \sigma(\omega,T) 
\frac{\cosh(\omega (\tau-1/(2 T)))}{\sinh(\omega/(2 T))}\, .
\label{corr}
\end{equation}
Detailed information on $G(\tau,T)$ could make it possible to
reconstruct the spectral function from the lattice
data. In practice, however, this turns out to be very difficult
task because the time extent is limited to $1/T$, see the discussion
in Ref.~\cite{Jakovac:2006sf} and references therein. 

\begin{figure}[htbp]
\begin{center}
\includegraphics[width=7.5cm]{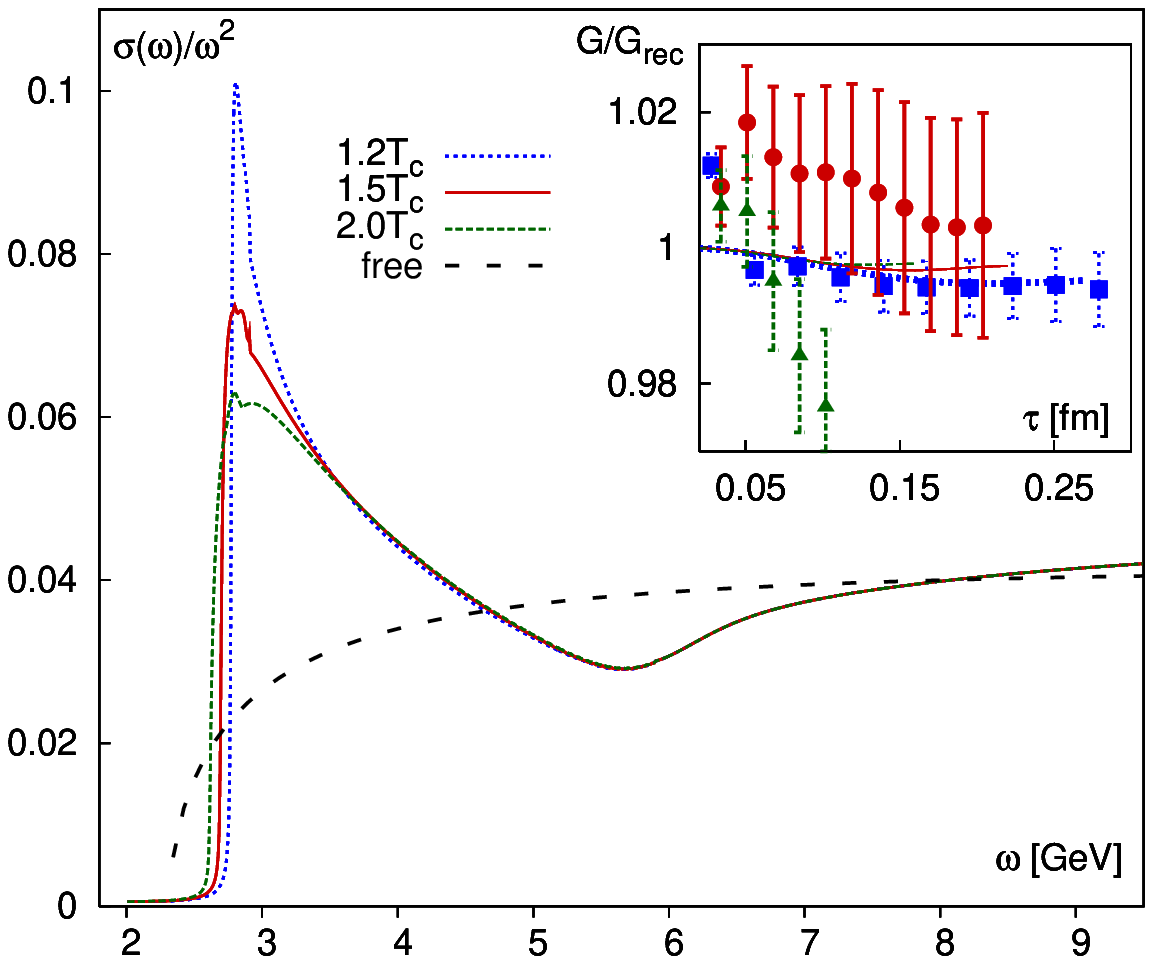}
\includegraphics[width=7.5cm]{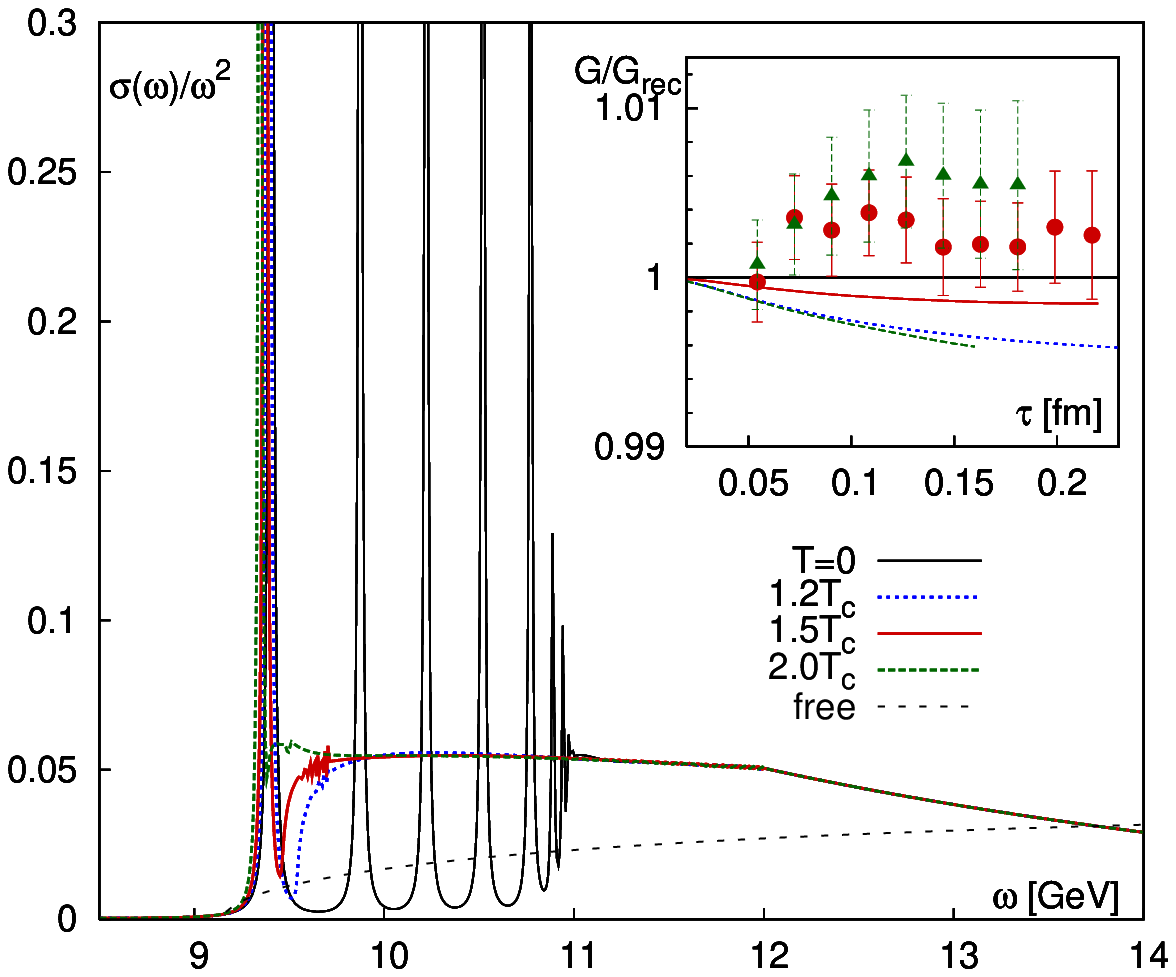}
\vspace*{-0.5cm}
\caption{The $S$-wave charmonium (left) and bottomonium (right) spectral 
functions calculated in potential models \cite{Mocsy:2007yj}. 
Insets: correlators compared to lattice data \cite{petr_hq08}.  The dotted curves are the
free spectral functions.}
\label{fig:spf} 
\end{center}
\vspace*{-0.3cm}
\end{figure}

The quarkonium spectral functions can be calculated in potential models 
using the singlet free energy from Fig.~\ref{fig:f1} or with different 
lattice-based potentials obtained using the singlet free energy
as an input  \cite{Mocsy:2007yj,Mocsy:2007jz}. 
The results of calculations in quenched QCD are shown in Fig.~\ref{fig:spf}
for $S$-wave charmonium and bottomonium  
spectral functions \cite{Mocsy:2007yj}.
All charmonium states
are dissolved in the deconfined phase, while the bottomonium $1S$
state may persist up to temperature of about $2T_c$. 
The temperature dependence of the Euclidean correlators can be predicted 
using Eq.~(\ref{corr}) and the calculated spectral functions. 
On the lattice the temperature dependence of the correlation function is
studied in terms of the ratio $G(\tau,T)/G_{rec}(\tau,T)$, where
\begin{equation}
G_{rec}(\tau,T)=\int_0^{\infty} d \omega \sigma(\omega,T=0)
\frac{\cosh(\omega (\tau-1/(2 T)))}{\sinh(\omega/(2 T))}\, .
\label{rec}
\end{equation}
If the spectral function is temperature independent $G/G_{rec}=1$.
It turns out, somewhat surprisingly, that the Euclidean correlation functions
in the pseudo-scalar channel 
show very little temperature dependence irrespective of whether a state 
remains bound (the $\eta_b(1S)$ )or not (the $\eta_c(1S)$). 
Note also that correlators from potential models are in accord with the 
lattice calculations (see insets in Fig.~\ref{fig:spf}). Initially, 
the weak temperature dependence of the correlators was considered to be 
evidence for the survival of different quarkonium
states \cite{Datta:2003ww}. It is now clear that this conclusion was premature. 
%\vspace*{-0.3cm}
%\vspace*{-0.3cm}
There is a large enhancement in the threshold region of the spectral functions
relative to the free spectral function, as shown in Fig.~\ref{fig:spf}.
This threshold enhancement compensates for the absence of bound states
and leads to Euclidean correlation functions with a very weak
temperature dependence \cite{Mocsy:2007yj}. It further indicates strong residual
correlations between the quark and anti-quark, even in absence of bound states.
Present lattice calculations of the spectral functions cannot discriminate between
bound state peaks and threshold enhancement.

Similar analysis has been done for the $P$-wave charmonium and bottomonium 
spectral functions \cite{Mocsy:2007yj,Mocsy:2007jz}. 
Here the contribution of the zero mode is important. The zero mode
contribution is responsible for large enhancement and strong temperature
dependence of $G/G_{rec}$ in the scalar and axial-vector channels 
\cite{Mocsy:2007yj,umeda07}. A smaller temperature dependence of $G/G_{rec}$
due to zero mode was observed in the vector channels as well \cite{derek06}.
Here the zero mode contribution is related to heavy quark diffusion.
A comprehensive study of the zero mode contribution was presented in 
Ref. \cite{petr_hq08} where it was concluded that almost the entire temperature
dependence of the correlation function is due to this contribution.
Furthermore, the zero mode contributions which are related to generalized
susceptibilities are well described by quasi-particle model \cite{petr_hq08}.
This provides additional input for in-medium properties of heavy quarks, e.g.
the temperature dependent heavy quark mass, which will be important for
constructing more realistic potential model at finite temperature.  
While the above studies have been performed in quenched approximation, 
recently attempts to extend
the quasi-particle model in the charm quark sector to full QCD have been presented in Ref. \cite{lat09vel}.   

To estimate the dissociation temperatures, i.e. the temperature above which no resonance like
peak are observed in the quarkonium spectral functions, it is not sufficient to consider
only the effects due to color screening.  One has to consider also the effect of thermal broadening
of different states. 
Upper bound on the
dissociation temperatures can be obtained from the analysis of the spectral 
functions calculated in potential model and using simple phenomenological estimates
for the thermal width \cite{Mocsy:2007jz}. This analysis shows that most of quarkonium states dissolve
at temperatures $<1.2T_c$ and only the ground state bottomonium may survive up to temperatures as high
as $2T_c$.

The application of potential models can be justified using an 
effective field theory approach.  The existence of distinct energy scales related to the 
heavy quark mass $m$, the inverse size $m v$ (where $v$ is the heavy quark 
velocity), and the binding energy $m v^2$ makes it possible
to construct a sequence of effective theories at zero temperature. 
The effective theory which emerges after integrating out the scales $m$ 
and $m v$ is pNRQCD.
The Lagrangian of this effective theory contains singlet and octet meson fields 
composed of heavy quarks, which are coupled to soft gluon fields at scale $m v^2$.
pNRQCD is equivalent to the potential model at $T=0$ at leading order,
where the coupling to the octet channel can be neglected.  
It is possible to extend this approach to finite
temperature where additional scales, the temperature $T$,
the Debye mass $m_D \sim g T$, and the magnetic scale $g^2 T$ are present.
In the weak coupling regime, $g \ll 1$, these scales are
well separated. Depending on how the thermal scales are
related to the zero temperature scales, the various hierarchies make it possible
to derive different effective theories for quarkonium bound states
at finite temperature \cite{Brambilla:2008cx}. In the weak-coupling QCD 
approach, thermal corrections to the potential are obtained when the
temperature is larger than the binding energy.  An important general result 
of these effective theories is that the potential acquires an imaginary part 
\cite{Laine:2006ns,Laine:2007qy,Laine:2007gj,Laine:2008cf,Brambilla:2008cx}.
The imaginary part of the potential smears out the bound state 
peaks of the quarkonium spectral function, leading to their dissolution prior
to the onset of Debye-screening in 
the real part of the potential (see {\it e.g.} the 
discussion in Ref.~\cite{Laine:2008cf}).

%%%%%%%%%%%%%%%%%%%%%%%%%%%
\section{Dynamical models for quarkonium production}

While knowing the quarkonium spectral functions in equilibrium QCD is 
necessary, it is insufficient to predict effects on their production in 
heavy-ion collisions because, unlike the light degrees of freedom, heavy 
quarks are not fully thermalized in heavy-ion collisions. Therefore, it is 
nontrivial to relate the finite temperature quarkonium spectral functions
to quarkonium production rates in heavy ion collisions without further model
assumptions. The bridge between the two is provided by dynamical models of 
the matter produced in heavy-ion collisions. Some of the simple models 
currently available are based on statistical recombination 
\cite{Andronic:2006ky}; statistical recombination and dissociation
\cite{Thews:2000rj,Zhao:2007hh}; or sequential melting \cite{Karsch:2005nk}. 
Here we highlight a more recent model, which makes closer contact with both
QCD and experimental observations \cite{Young:2008he}.

The bulk evolution of the matter produced in heavy-ion collisions is 
well modeled by hydrodynamics, see Ref.~\cite{Teaney:2009qa} for a recent 
review.  The large heavy quark mass makes it possible to model its interaction 
with the medium by Langevin dynamics \cite{Moore:2004tg}. Such an approach 
successfully  describes the anisotropic flow of charm quarks observed at RHIC 
\cite{Moore:2004tg,vanHees:2004gq,Gossiaux:2006yu} (see also Ref. \cite{Rapp:2009my} for
a recent review).  Potential models have shown that, in the absence of 
bound states, the $Q \overline Q$ pairs are correlated in space 
\cite{Mocsy:2007yj,Mocsy:2007jz}. This correlation can be modeled classically
using Langevin dynamics, including a drag force and a random force between the 
$Q$ (or $\overline Q$) and the medium as well as the forces between the $Q$ 
and $\overline Q$ described by the potential. It has been shown that a 
model combining an ideal hydrodynamic expansion of the medium with a
description of the correlated $Q \overline Q$ pair dynamics by the Langevin 
equation may explain 
why, despite the fact that a deconfined medium is created at RHIC, there is
only a $40-50\%$ suppression in the charmonium yield. The attractive potential
and the finite lifetime of the system prevents the complete decorrelation of
some of the $Q \overline Q$ pairs \cite{Young:2008he}. Once the system has
sufficiently cooled, these residual correlations make it possible for 
the $Q$ and $\overline Q$ to form a bound state. 
The details of the suppression pattern depend on the model parameters, such as $T_c$
and the way recombination is implemented. The later
appears to be important in the central collisions \cite{clint_new}. However, the results
are not very sensitive to the choice of the potential \cite{private}.

The above approach, which neglects quantum effects, is applicable only if 
there are no bound states, as it is likely to be the case for the $J/\psi$. If 
heavy quark bound states are present, as is probable for the $\Upsilon (1S)$, 
the thermal dissociation rate will be most relevant for 
understanding the quarkonium yield. It is expected that the interaction of a 
color singlet quarkonium state with the medium is much smaller than
that of heavy quarks. Thus, to first approximation, medium effects will only 
lead to quarkonium dissociation.

%%%%%%%%%%%%%%%
\section{Concluding remarks}

Lattice calculations of static quark anti-quark correlators provide
evidence for strong screening effects in the deconfined phase. 
Potential model calculations based on lattice QCD, as well as re-summed 
perturbative QCD calculations indicate that all charmonium states and 
excited bottomonium states dissolve in the deconfined medium. This leads to 
the reduction of the quarkonium production yield in heavy-ion collisions 
compared to binary-scaling of $pp$ collisions. Survival of residual
correlations in the deconfined phase as well as recombination effects
at lower temperatures lead to non-zero quarkonium yield in heavy ion 
collisions.

It turns out, that despite melting of quarkonium states corresponding
meson correlation function in Euclidean time show very little temperature
dependence. Almost the entire temperature dependence of the Euclidean 
correlation functions is due to the zero mode contribution. Statements 
about the existence of heavy quark bound states based on lattice correlation
functions and spectral functions need to be revisited. Yet, lattice 
results on correlation functions are still valuable tools for constraining
potential models. 

One of the great opportunities of the LHC and RHIC-II heavy-ion programs 
is measurement of bottomonium yields. From a theoretical perspective, 
bottomonium is an important and clean probe for at least two reasons. 
First, the effective field theory approach, which provides a link to first 
principle QCD, is more applicable for bottomonium due to better separation of 
scales and higher dissociation temperatures. Second, the heavier bottom quark 
mass reduces the importance of recombination effects, making bottomonium a 
good probe of dynamical models. 

%%%%%%%%%%%%%%%

\end{document}